\PassOptionsToClass{12pt}{revtex4-1}
\documentclass[manusript]{aastex631}
\usepackage[T1]{fontenc}
\usepackage[]{graphicx,tabularx,ctable,color,colortbl,amsmath,amsfonts,float,bm}
\usepackage[stixtwo]{fontsetup}

\shortauthors{Clarkson \& Kontar}

\begin{document}

\title{Instantaneous Spectral Bandwidth of Solar Type III Radio Bursts in the Interplanetary Medium}

\author[0000-0003-1967-5078]{Daniel L. Clarkson}
\affiliation{School of Physics \& Astronomy, University of Glasgow, Glasgow, G12 8QQ, UK}
\email{daniel.clarkson@glasgow.ac.uk}  

\author[0000-0002-8078-0902]{Eduard P. Kontar}
\affiliation{School of Physics \& Astronomy, University of Glasgow, Glasgow, G12 8QQ, UK}
\email{eduard.kontar@glasgow.ac.uk}  

\begin{abstract}
Magnetic reconnection during solar flares accelerates electron beams that generate type III radio bursts through the plasma emission process. Their instantaneous spectral bandwidth is commonly interpreted as a measure of the spatial extent of the emitting population, yet contributions from viewing angle and radio-wave propagation remain poorly constrained. We investigate bandwidths of 20 type III bursts observed by Parker Solar Probe between 0.15--6.5~MHz. The bandwidth increases with frequency and is consistent with previous measurements up to 100~MHz. For the first time, simultaneous observations from multiple spacecraft show similar bandwidths despite large differences in viewing geometry. We  further examine bandwidth contributions from intrinsic emission and propagation effects using numerical and analytical simulations, showing that the beam spatial extent is larger than required to explain observed bandwidths, while the spatial extent of Langmuir wave generation produces electromagnetic emission bandwidths near the observed lower limit. Temporal broadening due to radio-wave scattering can increase the bandwidth by causing emission from different frequencies to overlap in time. This effect depends strongly on turbulence level and becomes increasingly important for faster electron beams and narrower intrinsic emission profiles. The results demonstrate that observed type III bandwidths cannot be interpreted solely in terms of physically emitting regions: the intrinsic emission bandwidth follows the spatial extent of Langmuir wave generation, while propagation-induced temporal broadening provides an additional contribution that can reconcile intrinsic and observed bandwidths. Accounting for radio-wave propagation effects is essential when using type III burst bandwidths to infer the spatial properties of their emitting electron populations.
\end{abstract}


\section{Introduction}

Magnetic energy release in solar flares leads to the acceleration of electrons to non-thermal energies
\citep{2011SSRv..159..107H,2017LRSP...14....2B,2017ApJ...836...17A}. Escaping electron beams stream along the pervading magnetic field \citep{1998SSRv...86...61L,2017JSWSC...7A..30P,2023ApJ...958...18P,2026ApJS..286...18C}
routinely excite Langmuir waves that produce radio emission near the local plasma frequency $f_\mathrm{pe}$ and its harmonic $f_\mathrm{H}=2f_\mathrm{pe}$ via the plasma emission process  \citep{1958SvA.....2..653G,1985ARA&A..23..169D, 2017RvMPP...1....5M}. Radio emission is generated via the coalescence of Langmuir waves with ion sound waves or counter-propagating Langmuir waves. Given the decreasing plasma density $n_e$ with distance $r$ from the Sun where $f_\mathrm{pe}\,\mathrm{[MHz]}\simeq 8.93\times10^{-3}n_e(r)^{1/2}$, this process produces fast drifting structures in dynamic spectra---so-called type III bursts \citep{1950AuSRA...3..541W,1974SSRv...16..145F, 2014RAA....14..773R}. Therefore, the radio emission that forms the backbone of a type III burst is coupled with the spatial distribution of Langmuir waves. Type III bursts are a valuable diagnostic for both the environment through which they propagate \citep{2017NatCo...8.1515K, 2018ApJ...856...73C, 2023A&A...680A.106N, 2026ApJ...999..134C} and of the exciter itself \citep{1981ApJ...251..364L, 2007ApJ...663L.109K, 2011A&A...529A..66R, 2018A&A...614A..69R}.

Observationally, the bandwidth of type III bursts is not well analysed in the literature. \cite{1963ApJ...138..239H} measure the instantaneous bandwidth of 14 type III bursts by measuring the frequency range between the leading edge of the burst and its highest observed frequency. At 100~MHz, they find the average bandwidth is $\Delta{f}\simeq100$~MHz, associating this with an exciter length of 0.25~R$_\odot$. At frequencies between 10--30~MHz, \cite{2011SoPh..269..335M} measure average bandwidths of 10--20~MHz from 400 type III bursts using the half-power frequency widths in the dynamic spectra, finding the relation $\Delta f(f)\approx 0.6f$.
Between 30--70~MHz, \cite{2018A&A...614A..69R} use the Low Frequency Array to measure the bandwidth of 31 type III bursts, finding $\Delta{f}$ to vary between approximately 10--30~MHz, with a relation of $\Delta{f}(f)=0.44 - 0.0017f$, although the scatter in the measurements is large. To measure the bandwidth, they use the HWHM of the temporal profiles at each frequency to determine the rise and decay times, and then define the instantaneous bandwidth as the frequency range between the lowest frequency that has started emitting and the highest frequency that is still emitting. Recently, \cite{2024ApJ...972....2T} measure the FWHM $\Delta{f}$ using Gaussian fits between 2--12~MHz, finding values from approximately 2--10~MHz, with a relation of $\Delta{f}(f)=0.85f$.  Using a relation of $\Delta{f}(f)\approx 0.7f$ with a density model above an active region \citep{1958ApJ...128..664P, 1999A&A...348..614M}, one may estimate the characteristic spatial extent $\Delta{r}$ corresponding to the observed bandwidth at a given frequency. Since the plasma frequency scales as $f_\mathrm{pe}\propto n_e^{1/2}$, one finds that $\mathrm{d}f_\mathrm{pe}/\mathrm{d}r=(f_\mathrm{pe}/2n_e(r))(\mathrm{d}n_e/\mathrm{d}r)$. Thus, the spatial extent corresponding to an instantaneous bandwidth $\Delta f$ is $\Delta r\simeq 2(\Delta f/f_\mathrm{pe})L_n(r)$, where $L_n(r)=|\mathrm{d}\ln n_e(r)/\mathrm{d}r|^{-1}$ is the density scale height. At 1~MHz and 0.1~MHz, this gives $\Delta r_{1~\mathrm{MHz}}\simeq 5$~R$_\odot$ and $\Delta r_{0.1~\mathrm{MHz}}\simeq 40$~R$_\odot$, respectively.

The spatial distribution of Langmuir waves is governed by the evolution of the electron beam as it propagates through the corona and solar wind. \cite{1999SoPh..184..353M} showed that the electron beam propagates as a beam-plasma structure. The electron distribution undergoes quasi-linear relaxation, forming a plateau in velocity space as electrons with velocity $v \gg v_\mathrm{Te}$ near the front of the beam lose energy due to the interaction with Langmuir waves. Consequently, lower energy electrons near the tail of the beam absorb energy from Langmuir waves and are accelerated \citep{1972SoPh...24..444Z}.
Spatially nonuniform relaxation of the beam is expected to lead to the spatial broadening of both the electron beam and the region of Langmuir wave generation \citep{2024ApJ...976..233K}.
\cite{2014A&A...572A.111R} numerically modelled the evolution of an electron distribution function in time and space, finding that the frequency spread of Langmuir waves at a single location is small, giving $\Delta f/f = 0.06$. However, the simulated radio emission has bandwidths of $\Delta f/f = (0.4-0.6)$, which the authors attribute to the substantial spatial extent of the electron beam and the resulting spatial extent of Langmuir-wave generation.
\cite{2018ApJ...867..158R} investigate the spatial extent between the front and back of the exciting electron beam using numerical simulations, finding that the initial characteristics of the accelerated electron population affect the electron beam length---the width in frequency space is broader for higher beam densities and electron beams with lower initial spectral indices (since the beam length extends more due to stronger velocity dispersion). Thus, for relatively smooth decreases in density over large distances, the instantaneous bandwidth of type III bursts can be related to the spatial extent of the electron beam through the range of plasma frequencies between its front and back. For burst fine structures such as striae and spikes, the bandwidth can similarly provide information on the spatial extent of the emitting region, where density fluctuations localise the plasma emission to relatively narrow regions of space \citep[e.g.][]{2017NatCo...8.1515K, 2018ApJ...856...73C, 2021ApJ...917L..32C}.

Interpreting the observed bandwidths in terms of the spatial extent of the emitting region implicitly assumes that the observed emission directly reflects the properties of the source region. However, radio waves are strongly affected by propagation through an inhomogeneous plasma. Indeed, the turbulent nature of the corona and heliosphere leads to small angular scattering of the emission, and numerous works have shown that the scattering rate is higher perpendicular to the local magnetic field due to anisotropy $\alpha=q_\parallel/q_\perp$ in the density turbulence spectrum $\mathbf{q}$ \citep[e.g.][]{2019ApJ...884..122K, 2023ApJ...956..112K}, with typical values between $\alpha=(0.1-0.4)$ \citep{2017NatCo...8.1515K, 2020ApJ...905...43C, 2020ApJ...898...94K, 2021ApJ...917L..32C, 2021A&A...656A..34M, 2023ApJ...956..112K}. Scattering off small-scale density inhomogeneities does not affect the bandwidth of the radio emission significantly; in fact, radio wave scattering in the corona is often considered to be elastic such that the frequency is conserved \citep[e.g.][]{2019ApJ...884..122K}, and whilst a small degree of broadening in frequency can occur, this is significantly smaller (widths measured in hertz) than the observed burst bandwidth \citep[see][for the effect of inelastic scattering on spacecraft signals observed through the solar atmosphere]{2024ApJ...968...72A}. However, attributing the observed burst bandwidth solely to the spatial extent of the emitting region neglects the influence of temporal broadening during radio-wave propagation. \cite{2023ApJ...956..112K} use numerical simulations of anisotropic radio-wave scattering to show that radio burst time profiles are dominated by propagation effects. 
For type III bursts, this is particularly evident for emission below a few megahertz. If the temporal profile is sufficiently broadened, emission produced at different times can overlap across adjacent frequency channels, leading to an apparent increase in instantaneous bandwidth. Consequently, interpreting such broadened bandwidths without accounting for scattering may lead to an overestimation of the intrinsic spatial emission region. Moreover, the simultaneous observation of type III bursts by multiple spacecraft at different helio-longitudes provides a unique opportunity to investigate the influence of viewing angle on the measured bandwidth. 

In this study, we analyse the bandwidths of kilometric type III radio bursts observed by spacecraft between 0.15~MHz and 6.5~MHz (Section \ref{sec:obs}), and use simulations of type III bursts to investigate the impact of temporal broadening due to radio-wave propagation effects on instantaneous bandwidth measurements (Sections \ref{sec:numerical_sim} and \ref{sec:analyical_sim}). Section \ref{sec:discussion_conclusion} presents the discussion and conclusion.

\section{Spectral Width of Observed Type III Bursts}\label{sec:obs}

\begin{figure}
    \centering
    \includegraphics[width=0.6\linewidth]{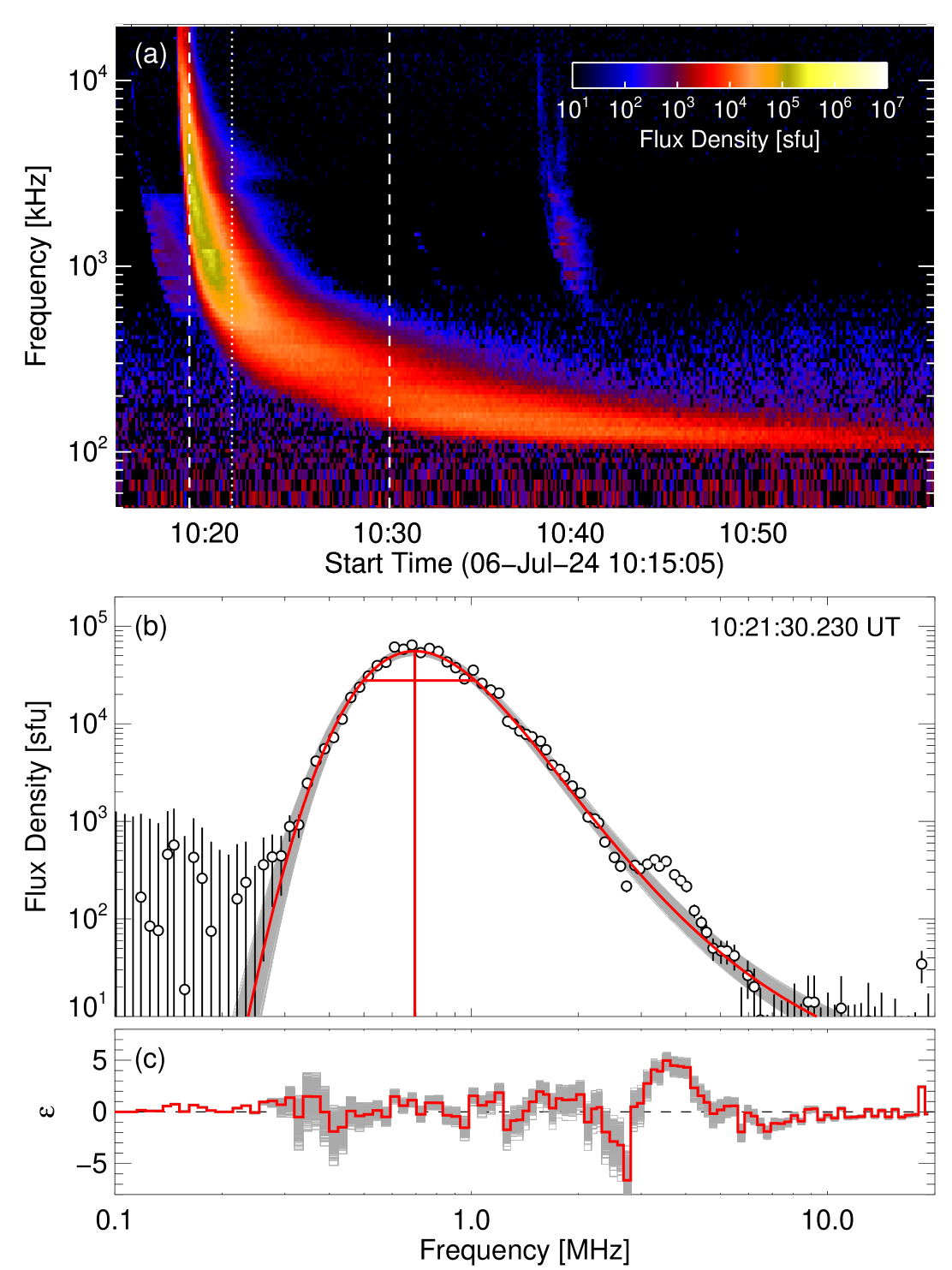}
    \caption{(a) Example type III burst observed by PSP on 2024 July 6 near 10:20 UT when PSP was approximately 62.5~R$_\odot$ from the Sun (near $f_\mathrm{pe}\simeq100$~kHz). The vertical dashed lines bound the times where the instantaneous bandwidth is measured. The dotted line shows the timing of the intensity profile in the panel (b). (b) Example fit (red curve) to the observed $I(f)$ profile (open circles) using equation (\ref{eq:If}). The vertical and horizontal red lines show the fitted $\Delta{f}$ and the frequency of $f_0$. The grey region shows the collated fits from Monte-Carlo sampling used to define the uncertainty in $\Delta{f}$ and $f_0$. (c) Normalised residuals $\epsilon_i=(I(f_i)-I_\mathrm{fit}(f_i))/\delta I(f_i)$. The grey regions shows the collated residuals from the Monte-Carlo sampling, and the red curve corresponds to the red fit in panel (b).}
    \label{fig:ds_eg}
\end{figure}

To measure the frequency bandwidth of kilometric type III radio bursts, we use observations from the Parker Solar Probe \citep[PSP;][]{2016SSRv..204....7F} FIELDS/RFS instrument \citep{2016SSRv..204...49B, 2017JGRA..122.2836P}. An example type III burst observed by PSP is shown in Figure \ref{fig:ds_eg}. The instantaneous bandwidth profiles of the type III bursts are asymmetric, with a shorter rise than decay---i.e. the segment of the intensity profile at the lower frequencies below the peak drops off faster towards background levels. 
We use an asymmetrical function of the form:
\begin{equation}\label{eq:If}
    I_\mathrm{fit}(f) = I_0\exp{\left[-\left(\frac{f-f_0}{a+bf}\right)^2\right]},   
\end{equation}
to fit the observed intensity profiles $I(f)$ as a function of frequency at a given time. Here, $I_0$ and $f_0$, $a$, and $b$ are free parameters of the fit. $I_0$ and $f_0$ give the peak flux density and peak frequency, whilst $a$ and $b$ are used to determine the rise and decay, as used for asymmetric time profiles by \cite{2025ApJ...978...73C}. We fit the instantaneous intensity profiles only during times when the burst was still emitting in the lowest frequency channel used in the fit, as shown in Figure \ref{fig:ds_eg}. This also ensures that no measurement is taken beyond times where the burst emission is not observed due to the local plasma frequency at the spacecraft location. Prior to fitting, the background is subtracted from each frequency-flux profile, where the background $I_\mathrm{bg}$ for each frequency channel is defined as the average flux density within a quiet time period before or after each burst.
The uncertainty on the flux density $\delta I$ is a combination of the burst intensity and the background level with a 12\% signal error given as $\delta I(f)=\sqrt{\left(0.12I(f)\right)^2 + \left(0.12I_\mathrm{bg}\right)^2}$ (see the discussion of the uncertainties and background in \citep{2025NatSR..1511335C}). Typically, the type III burst intensity is much larger than the background, and so the bandwidth is not sensitive to the background selection Figure \ref{fig:ds_eg}.

\begin{figure}[htb!]
    \centering
    \includegraphics[width=.49\linewidth]{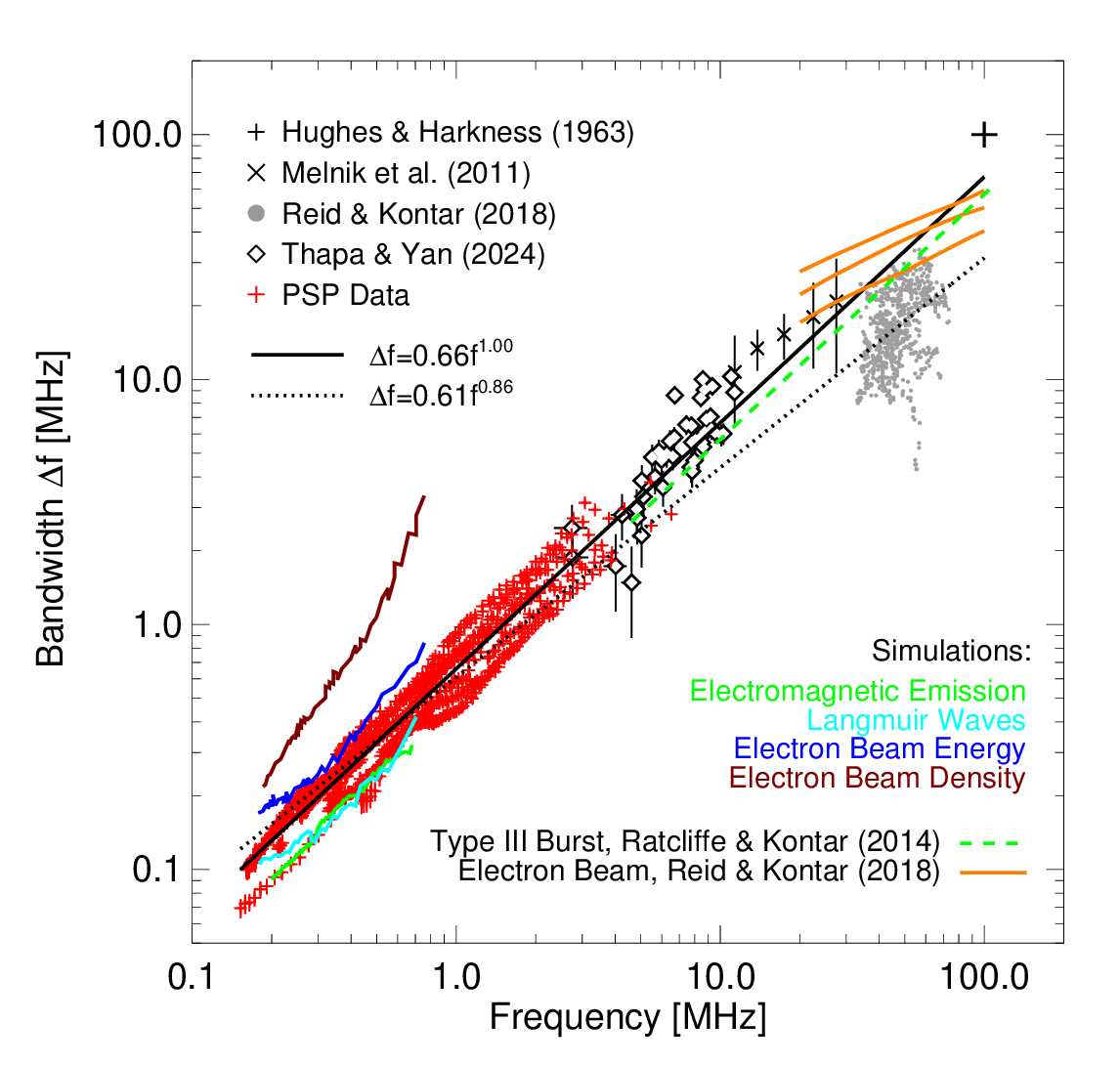}
    \includegraphics[width=.49\linewidth]{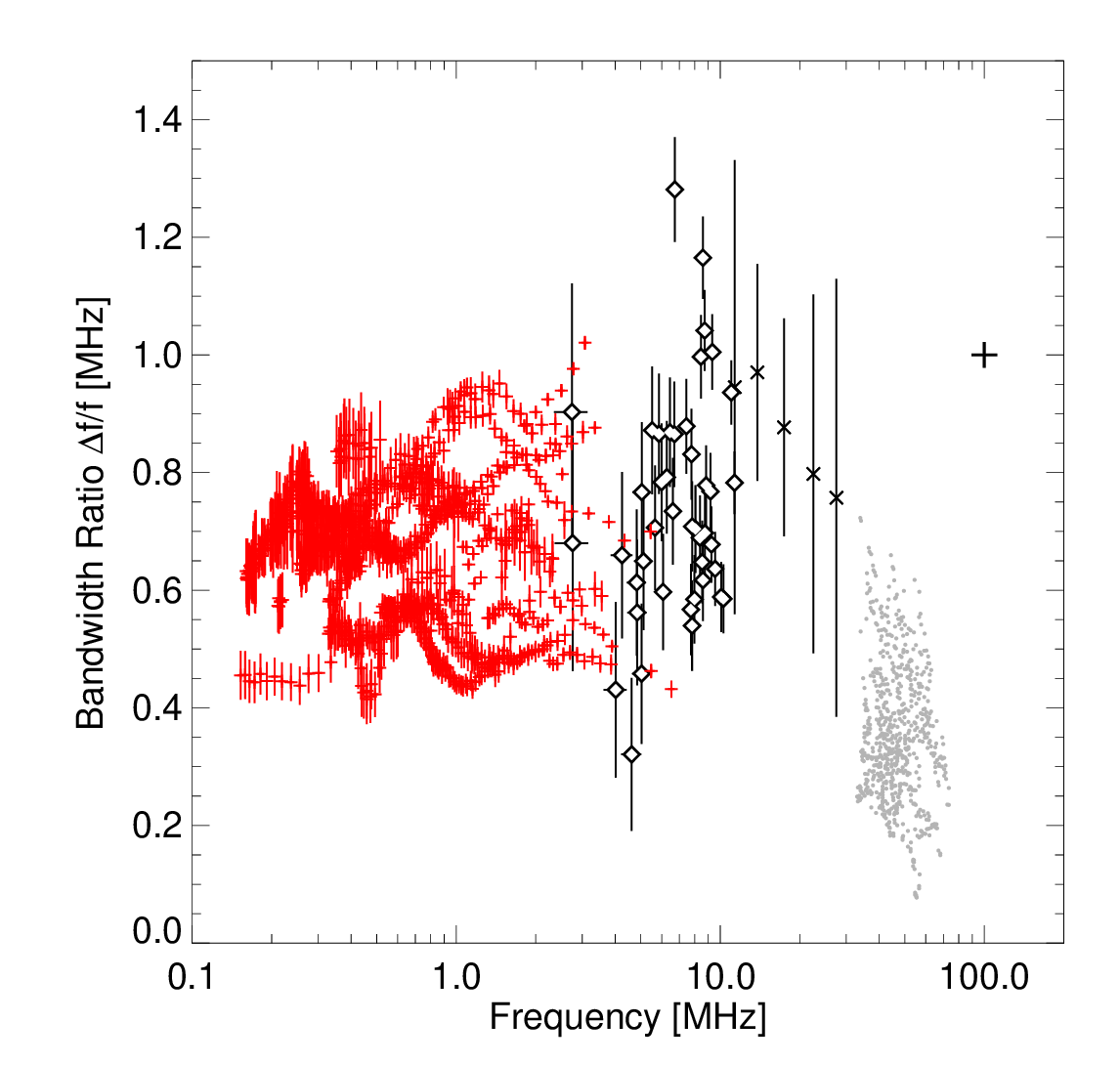}  
    \caption{Type III burst bandwidth $\Delta{f}$ (left) and bandwidth ratio $\Delta{f}/f$ (right) measured from \cite{1963ApJ...138..239H, 2011SoPh..269..335M, 2018A&A...614A..69R, 2024ApJ...972....2T}, and from 20 PSP observations (red) fitted using equation \ref{eq:If} at the FWHM level. We note that \cite{1963ApJ...138..239H} do not define an intensity level at which the bandwidth is measured, and \cite{2011SoPh..269..335M} and \cite{2024ApJ...972....2T} measure the FWHM. The dotted line shows a fit to the full collated data set, and the solid line shows a fit to the data excluding that from \cite{2018A&A...614A..69R}. Panel (a) also includes the FWHM spectral widths of the electron beam determined by the spatial spread in beam density (brown) and beam energy (blue), along with the spectral width of Langmuir wave energy (light blue), and electromagnetic emission (green) from the numerical simulation described in section \ref{sec:numerical_sim}. The spatial widths and peak position are converted to bandwidth and frequency using the density model of equation \ref{eq:n}. The dashed green line shows the bandwidth of simulated type III bursts from Figure 5 of \cite{2014A&A...572A.111R}. The orange lines show numerically simulated electron beam widths from Figure 14(a) of \cite{2018ApJ...867..158R}. The upper, middle, and lower lines represent different initial velocity spectral indices of 7, 8, and 9 respectively.}
    \label{fig:df_obs}
\end{figure}

Figure \ref{fig:ds_eg}(b) shows an example $I(f)$ profile at 10:17:25.6 UT from the burst shown in panel (a). The fitted function characterises the peak of the data near $f_0=0.69$~MHz with a FWHM bandwidth of $\Delta{f}=\left(0.53\pm0.01\right)$~MHz. The uncertainty on $\Delta{f}$ was estimated via Monte Carlo sampling. We generated 1000 realisations of the fitted parameters by drawing from Gaussian distributions defined by their $1\sigma$ uncertainties, recomputed the model for each realisation, and re-measured $\Delta{f}$. The standard deviation of the resulting distribution was adopted as the error on $\Delta{f}$. Figure \ref{fig:df_obs} presents the measured values of $\Delta{f}(f)$ across 20 type III bursts. The data are plotted against the peak frequency of each fit, and combined with measurements from other studies up to 100~MHz. The lower bandwidths reported by \cite{2018A&A...614A..69R} may be attributable to their different bandwidth definition, which is based on the half-width at half-maximum (HWHM) timing boundaries of the burst rather than an instantaneous spectral bandwidth measure at a given level. Including this data in a power-law fit to all measurements gives a relation of $\Delta{f}(f)=0.61f^{0.86}$, whilst excluding it gives $\Delta{f}(f)=0.66f^{1.0}$.

\subsection{Multi-Spacecraft Bandwidth Measurements}

\begin{figure}[htb!]
    \centering
    \includegraphics[width=0.49\linewidth]{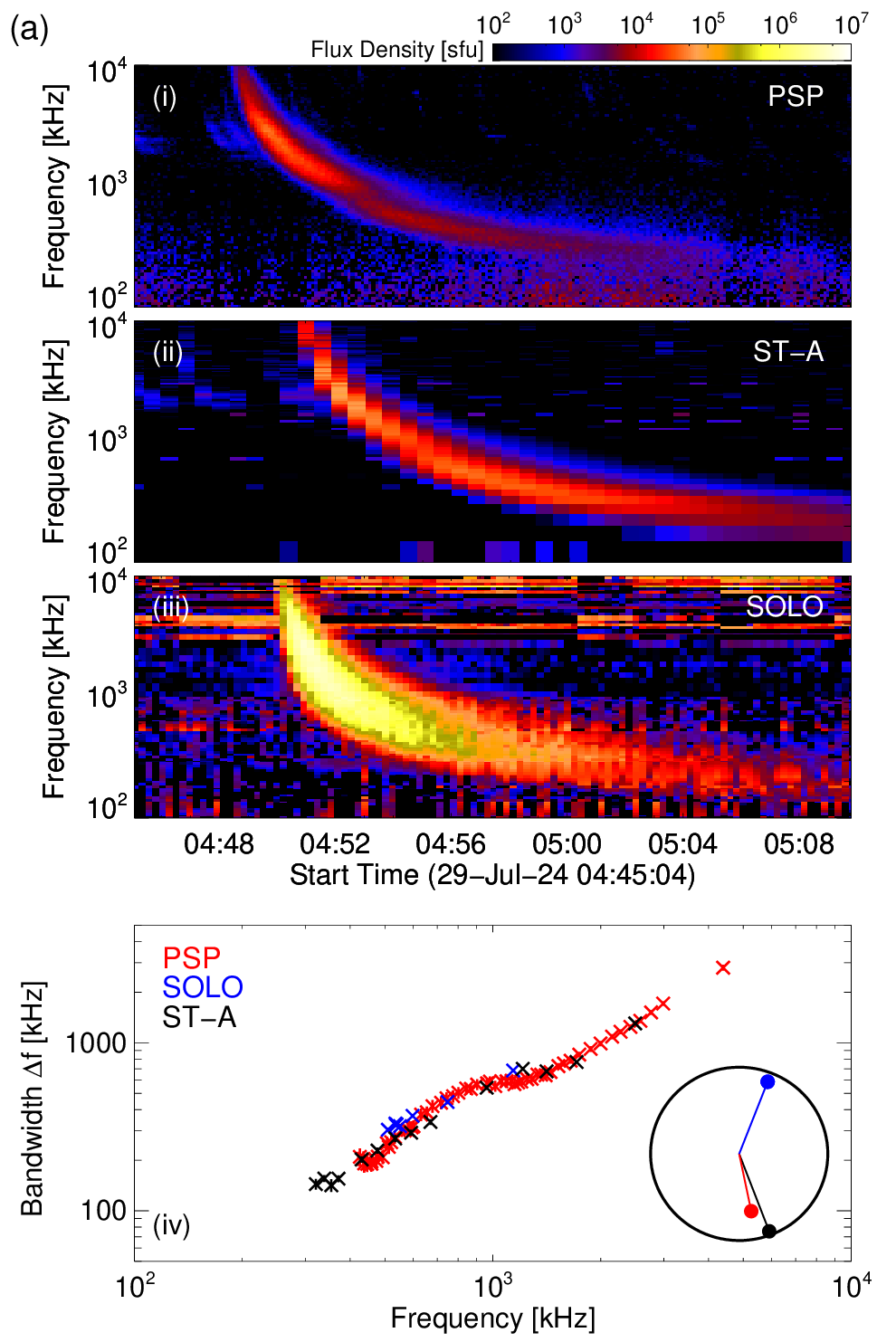}
    \includegraphics[width=0.49\linewidth]{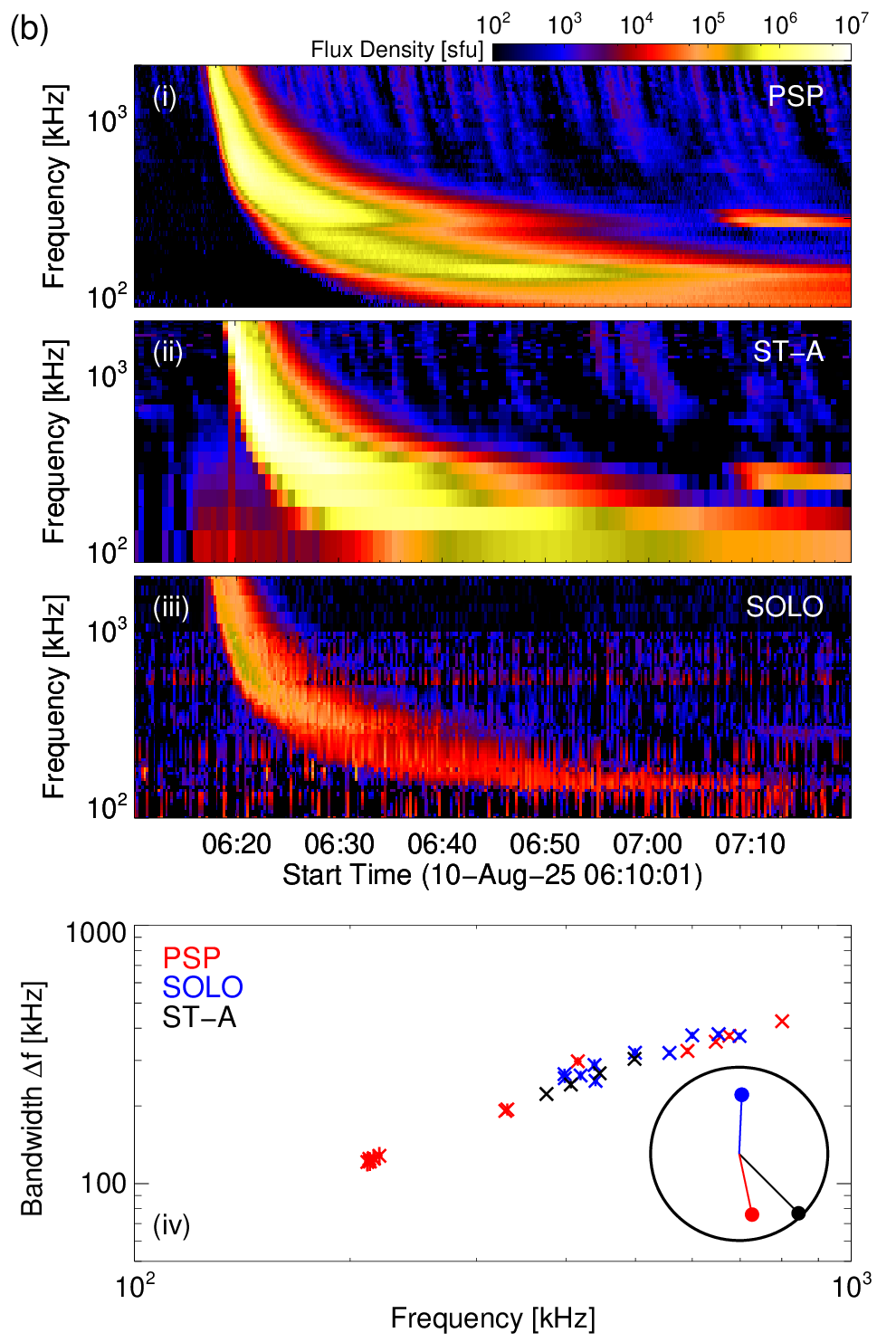}
    \caption{Type III bursts observed by multiple spacecraft on (a) 2024-July-29 near 04:50 UT and (b) 2025-Aug-10 near 06:20, along with the associated instantaneous bandwidth measurements. (i)--(iii) Dynamic spectra observed by PSP, ST-A, and SolO. (iv) Instantaneous bandwidth as a function of central frequency. The insets of panel (iv) show the relative spacecraft heliocentric locations. In column (a), the heliolongitudes are $11.6\arcdeg$ (PSP), $20.8\arcdeg$ (ST-A), and $158.9\arcdeg$ (SolO). For column (b), the heliolongitudes are $11.7\arcdeg$ (PSP), $44.3\arcdeg$ (ST-A), and $177.6\arcdeg$ (SolO). The outer boundary represents 1~au.}
    \label{fig:multi_sc_20240729}
\end{figure}

Figure \ref{fig:multi_sc_20240729} shows two type III burst events observed simultaneously by multiple spacecraft. In column (a), PSP and STEREO-A \citep[ST-A;][]{2008SSRv..136..487B} share approximately the same longitude (separated by $10\arcdeg$), whilst Solar Orbiter \citep[SolO;][]{2020A&A...642A...1M, 2020A&A...642A..12M} is separated from PSP by $147\arcdeg$ and observed the brightest emission, suggesting that the source was closer in longitude to SolO. In column (b), PSP and SolO are separated by $166\arcdeg$, whilst ST-A is $33\arcdeg$ ahead of PSP. Here, SolO received the faintest emission, indicating that the source was closer to in longitude to both PSP and ST-A. In each column, panels (iv) show that the bandwidth from all spacecraft are effectively identical at all comparable frequencies, despite drastically different helio-longitudes, and therefore observing angles with respect to the emission location.

%\begin{figure}[htb!]
%    \centering
%    \includegraphics[width=0.49\linewidth]{figures/multi_sc/20240505/ds.jpg}
%    \includegraphics[width=0.49\linewidth]{figures/multi_sc/20240505/sc_pos.jpg}\\
%    \includegraphics[width=0.8\linewidth]{figures/multi_sc/20240505/df_multi_sc.jpg}
%    \caption{Type III burst observed by multiple spacecraft on 2024-May-5 near 14:00 UT, and the associated instantaneous bandwidth measurements. (a) Dynamic spectra observed by PSP, ST-A, and SolO. (b) Spacecraft locations in HEE coordinates. (c) Instantaneous bandwidth as a function of central frequency.}
%    \label{fig:multi_sc_20240505}
%\end{figure}

%\begin{figure}[htb!]
%    \centering
%    \includegraphics[width=0.49\linewidth]{figures/multi_sc/20240727/ds.jpg}
%    \includegraphics[width=0.49\linewidth]{figures/multi_sc/20240727/sc_pos.jpg}\\
%    \includegraphics[width=0.8\linewidth]{figures/multi_sc/20240727/df_multi_sc.jpg}
%    \caption{As in Figure \ref{fig:multi_sc_20240505} for a burst observed on 2024-July-27 near 00:30 UT.}
%    \label{fig:multi_sc_20240727}
%\end{figure}

\section{Spectral Width of Simulated Type III Bursts}

We consider the observed bandwidth of type III bursts to consist of two contributions: the intrinsic emitting spatial length $\Delta{f}_\mathrm{int}$, and propagation-induced broadening, $\Delta{f}_\mathrm{scat}$. As an approximation, we take these contributions to be combined in quadrature as
\begin{equation}\label{eq:df}
    \Delta{f}^2\simeq\Delta{f}_\mathrm{int}^2 + \Delta{f}_\mathrm{scat}^2 \simeq \left(\left|\frac{\mathrm{d}f}{\mathrm{d}r}\right|\Delta{r}\right)^2 + \Delta{f^2_\mathrm{scat}},
\end{equation}
where $\Delta{r}=v_\mathrm{b} t_\mathrm{e}$ is the emitting spatial length determined by the exciter velocity $v_\mathrm{b}$ and the intrinsic emission time $t_\mathrm{e}$.
Here, we explore each contribution through numerical and analytical simulations of type III radio bursts. Through figures (\ref{fig:electrons_Lwaves_EM_eg})--(\ref{fig:beam_Lexp_comparison}), we set the emission mode to the fundamental level such that the emitted frequency $f$ corresponds approximately to the local plasma frequency, ($f\simeq f_{\mathrm{pe}}$). In Appendix figure \ref{fig:beam_turbScaling_comparison_harmonic}, we explore the implications of a harmonic emission mode.

\subsection{Numerical Simulation of Electron Beam Evolution \& Langmuir Wave Growth}\label{sec:numerical_sim}

To show the extent of the bandwidth contribution from the intrinsic electron beam and plasma emission processes, we first use numerical simulations of the evolution of an electron beam distribution $f(v,r,t)$~[cm$^{-4}$ s] and the corresponding growth of Langmuir waves measured by the spectral energy density $W_\mathrm{L}(v,r,t)$~[erg cm$^{-2}$]. Their evolution is computed using the parameters set in \cite{2026ApJ...999..134C} (see their section 4) using the code by \citet{2001SoPh..202..131K, 2001CoPhC.138..222K}. The code solves one-dimensional kinetic equations:
    \begin{equation}\label{eq:Frvt}
            \frac{\partial{f}}{\partial{t}} + v\frac{\partial{f}}{\partial{r}} = \frac{4\pi^2 e^2}{m_\mathrm{e}^2} \frac{\partial}{\partial{v}}\left(\frac{W_\mathrm{L}}{v}\frac{\partial{f}}{\partial{v}}\right)\;,
    \end{equation}
    \begin{equation}\label{eq:Wrvt}
            \frac{\partial{W_\mathrm{L}}}{\partial{t}} + \frac{\partial{\omega}}{\partial{k_\mathrm{L}}}\frac{\partial{W_\mathrm{L}}}{\partial{r}} - \frac{\partial{\omega_\mathrm{pe}}}{\partial{r}}\frac{\partial{W_\mathrm{L}}}{\partial{k_\mathrm{L}}} = \frac{\pi\omega_\mathrm{pe}}{n_\mathrm{e}}v^2 W \frac{\partial{f}}{\partial{v}} - \gamma W_\mathrm{L},\;\;\; \omega_\mathrm{pe}=k_\mathrm{L}v,
    \end{equation}
where $r$ is the heliocentric distance, $m_e$ and $e$ are the electron mass and charge, $\omega_\mathrm{pe}$ is the plasma frequency, and $k_\mathrm{L}$ is the wavenumber of the Langmuir waves. The term $\gamma W_\mathrm{L}$ represents the Landau damping of the plasma waves, with $\gamma=\pi n_e e^4 z_p^2 \ln\Lambda /m_\mathrm{e}^2 v_\mathrm{Te}^3$ where $\ln\Lambda\approx20$ is the Coulomb logarithm, $z_p=1.18$ gives the average atomic number in the photosphere \citep{2011A&A...536A..93J}, $v_\mathrm{Te}=\sqrt{k_\mathrm{B}T/m_\mathrm{e}}$ is the thermal velocity for temperature $T$, and $k_\mathrm{B}$ is the Boltzmann constant. Equations \ref{eq:Frvt} and \ref{eq:Wrvt} interact resonantly under the condition $\omega_\mathrm{pe}=k_\mathrm{L}v$, on the quasi-linear timescale $\tau_\mathrm{ql}(r)=n_e(r)/(n_\mathrm{b}\omega_\mathrm{pe}(r))$.
The initial distribution of electrons is given by
\begin{equation}
        f(v,r,t=0) = \frac{2n_\mathrm{b}}{v_0}\exp{\left(-\frac{r^2}{d^2}\right)},\,\, 0\leq v\leq v_0
\end{equation}
which forms an initial plateau in velocity space from $v=0$ to an upper limit of $v=v_0$. Here, $n_\mathrm{b}=2\times10^{-3}$~cm$^{-3}$ is the beam density, $d=2\times10^{10}$~cm is the spatial size, and $v_0=3\times10^9$~cm s$^{-1}\simeq0.1c$. At $t=0$, the distribution is centred on $r_0=1.1\times$10$^{11}$~cm ($2.6$~R$_\odot$). The background spectral energy density is set to $W_\mathrm{L}(r,v,t=0) = 10^{-8} \approx m_\mathrm{e}n_\mathrm{b}v_\mathrm{b}^3/\omega_\mathrm{pe}$, and the background plasma density is given by \citep{2019ApJ...884..122K}:
\begin{equation}\label{eq:n}
    n_e(r)=4.8\times10^9\left(\frac{r}{R_\odot}\right)^{-14} + 3\times10^8\left(\frac{r}{R_\odot}\right)^{-6} + 1.39\times10^6\left(\frac{r}{R_\odot}\right)^{-2.3}\,\mathrm{[cm^{-3}]}.
\end{equation}

\begin{figure}[htb!]
    \centering
    \includegraphics[width=.6\linewidth]{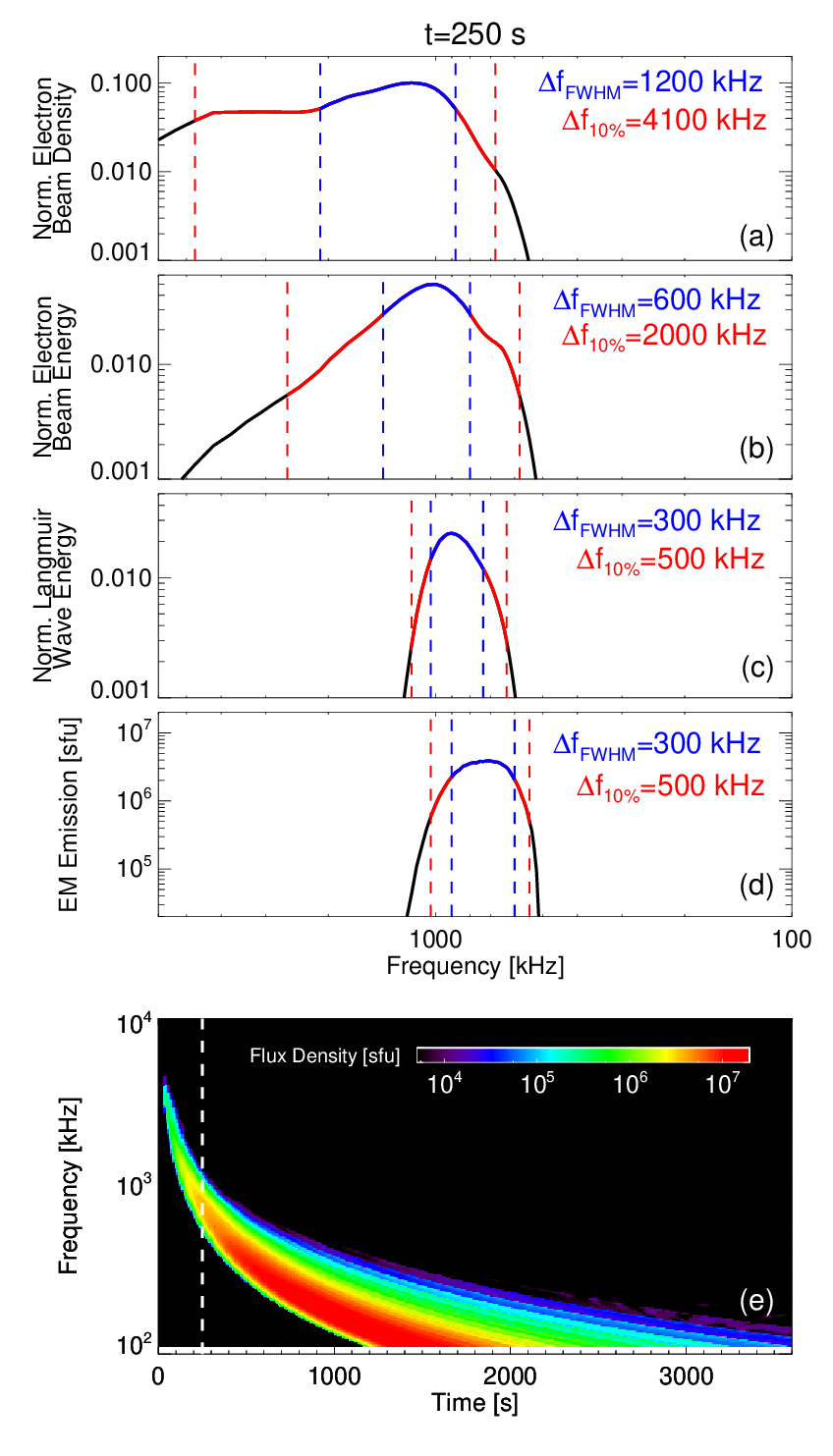}
    \caption{Numerical simulation results showing the normalised (a) electron beam density, (b) electron beam energy, (c) Langmuir wave energy, and (d) electromagnetic emission as a function of frequency $\sim250$~s after injection. The blue curves highlight the FWHM, bounded by the blue dashed lines and the red curves highlight the 10\% level. Note that the tail of the electron beam at higher densities has lower energy that does not produce Langmuir waves and radio emission. Panel (e) shows the simulated dynamic spectrum.}
    \label{fig:electrons_Lwaves_EM_eg}
\end{figure}

Figure \ref{fig:electrons_Lwaves_EM_eg} presents the normalised electron energy and normalised Langmuir wave spectral energy density as a function of distance at $\sim250$~s after injection, integrated over all electron velocities. The blue segments of each curve indicate the FWHM. The electron beam density extends over a broad frequency range, reflecting the range of electron velocities at injection, from zero to the maximum beam velocity $v_0$. The slower electrons travel relatively short distances during propagation and remain closer to the injection site, while the faster electrons propagate progressively farther ahead, producing an extended spatial distribution of the beam. The spatial extent of the electron distribution is further modified by the redistribution of electron energy through interactions with Langmuir waves, which alters the velocity distribution of resonant electrons as the beam propagates. The Langmuir wave energy density is concentrated at the leading edge of the electron beam, where the higher velocity electrons are located. This reflects the preferential generation of Langmuir waves by the faster electrons at the leading edge of the beam, which outrun the slower electrons during propagation and create a positive velocity-space gradient that drives the bump-on-tail instability. Beam-driven Langmuir-wave growth results in $W_\mathrm{L} \gg W_\mathrm{S}$, where $W_\mathrm{L}$ and $W_\mathrm{S}$ are the spectral energy densities of Langmuir waves and ion-sound waves, respectively. The non-linear decay process $L \rightarrow T + S$, where $L$, $T$, and $S$ denote Langmuir waves, transverse electromagnetic waves, and ion-sound waves, respectively, causes exponential growth of ion-sound waves. This increases $W_\mathrm{S}$ and the electromagnetic emission brightness temperature $T_b$ until $W_\mathrm{S} \gg W_\mathrm{L}$, at which point the emission process saturates at the level given by \citep{2017SoPh..292..117L, 2018ApJ...867..158R, 2021NatAs...5..796R}:
\begin{equation}
    T_b(v,r,t) \approx \frac{(2\pi)^2}{k_B}\frac{v^2}{\omega_\mathrm{pe}^2} W_\mathrm{L}(v,r,t).
\end{equation}
Taking the peak value of $T_b(v,r,t)$ at each distance for a given time provides the brightness temperature as a function of distance $T_b(r,t)$, which is then mapped to frequency using equation \ref{eq:n} to construct the dynamic spectrum $T_b(f,t)$. We convert the brightness temperature to a flux density in solar flux units (sfu; where $1\,\mathrm{sfu}=10^{-19}\,\mathrm{erg\,s^{-1}\,cm^{-2}\,Hz^{-1}}$) using apparent source sizes $\theta_\mathrm{FWHM}(f)$ as observed at 1~au as a function of frequency from the radio-wave propagation simulations of \cite{2023ApJ...956..112K} (see their Figure 4). In the Rayleigh--Jeans limit, the corresponding flux density is
\begin{equation}\label{eq:S}
    S(f,t)
    \simeq
    \left(\frac{2k_Bf^2}{c^2}\right)
    T_b(f,t)
    \left(\frac{\pi}{4\ln2}\right)
    \theta_{\rm FWHM}^2(f),
\end{equation}
where $T_b(f,t)$ is treated as the peak brightness temperature and the factor $\left(\frac{\pi}{4\ln2}\right)\theta_{\rm FWHM}^2(f)$ accounts for the integration of the assumed circular two-dimensional Gaussian source over solid angle. The resulting dynamic spectrum is shown in Figure \ref{fig:electrons_Lwaves_EM_eg}(d). We further note that the bandwidth of the electromagnetic emission is identical whether measured in brightness temperature or flux density, since $\theta_\mathrm{FWHM}(f)\propto 1/f$ \citep{2023ApJ...956..112K} removes the frequency dependence from Equation \ref{eq:S}, thereby preserving the spectral shape.

The bandwidth measured from the type III dynamic spectrum closely follows the Langmuir wave energy density width (Figure \ref{fig:electrons_Lwaves_EM_eg}). This is expected because the radio emission in the model is calculated directly from the Langmuir wave spectrum. Both quantities at the FWHM level are smaller by a factor of 4 than the width of the electron beam in frequency space. The Langmuir wave energy density bandwidth also depends on the initial electron beam width. A spatially smaller beam will generate Langmuir waves over a smaller region of space, resulting in a type III burst with a narrower bandwidth. In the simulation presented here, the bandwidth of the type III burst with typical interplanetary electron beam parameters lies at the lower boundary of the observed bandwidths from PSP, whereas the width of the electron beam in frequency space is too wide to explain the observations. This indicates that type III bandwidths are more closely associated with the spatial extent of Langmuir wave generation than with the full extent of the electron beam. Accordingly, in Equation \ref{eq:df}, we assume $\Delta f_\mathrm{int}\approx\Delta f_\mathrm{LW}$. In the next section, we explore the additional contribution to the observed bandwidths given by $\Delta f_\mathrm{scat}$.

\subsection{Analytical Intrinsic \& Scattered Type III Bursts}\label{sec:analyical_sim}

To explore the influence of broadening of the burst time profile on the instantaneous bandwidth, we use analytical simulations of a propagating electron beam emitting radio waves over time with varied beam velocities, widths, and expansion rates, as well as varied turbulence scaling factors. Below $\sim1$~MHz, the numerical type III burst presented in section \ref{sec:numerical_sim} becomes asymmetric in time, independent of propagation effects. As such, we model the intrinsic burst $I_\mathrm{int}(r,t)$ at each distance (frequency) channel as an asymmetric Gaussian:
\begin{equation}\label{eq:analytical_typeIII}
    I_\mathrm{int}(r,t) = A(r)w_i(f(r))
    \begin{cases}
        \displaystyle
        \exp\!\left(-\frac{(t - r/v_0)^2}{2\sigma_\mathrm{r}^2}\right), & t \leq r/v_0 \\
        \displaystyle
        \exp\!\left(-\frac{(t - r/v_0)^2}{2\sigma_\mathrm{d}^2(r)}\right), & t > r/v_0.
    \end{cases}
\end{equation}
Here, $r/v_0$ is the peak arrival time of the electron beam with speed $v_0$ at radial distance $r$. The leading edge is described by a Gaussian whose temporal standard deviation is determined by the initial beam length $d$ as $\sigma_\mathrm{r}=d/v_0$. The leading edge is assumed to be defined by the fastest electrons in the distribution, which arrive first and maintain a relatively sharp onset. To represent the effects of continual velocity dispersion, whereby slower electrons progressively lag behind the fastest electrons, the trailing edge is allowed to broaden with propagation distance through $d_\mathrm{exp}(r)=d\left[1 + (r/L)^2\right]$, giving a distance-dependent temporal width $\sigma_\mathrm{d}(r)=d_\mathrm{exp}(r)/v_0$. The prefactor $A(r)=\left[\sqrt{\pi/2}\left(\sigma_\mathrm{r} + \sigma_\mathrm{d}\right)\right]^{-1}$ normalises the asymmetric profile such that its integral remains unity. The spatial position of the exciter maps to frequency using the density model of equation \ref{eq:n}.
\begin{figure}[htb!]
    \centering
    \includegraphics[width=1\linewidth]{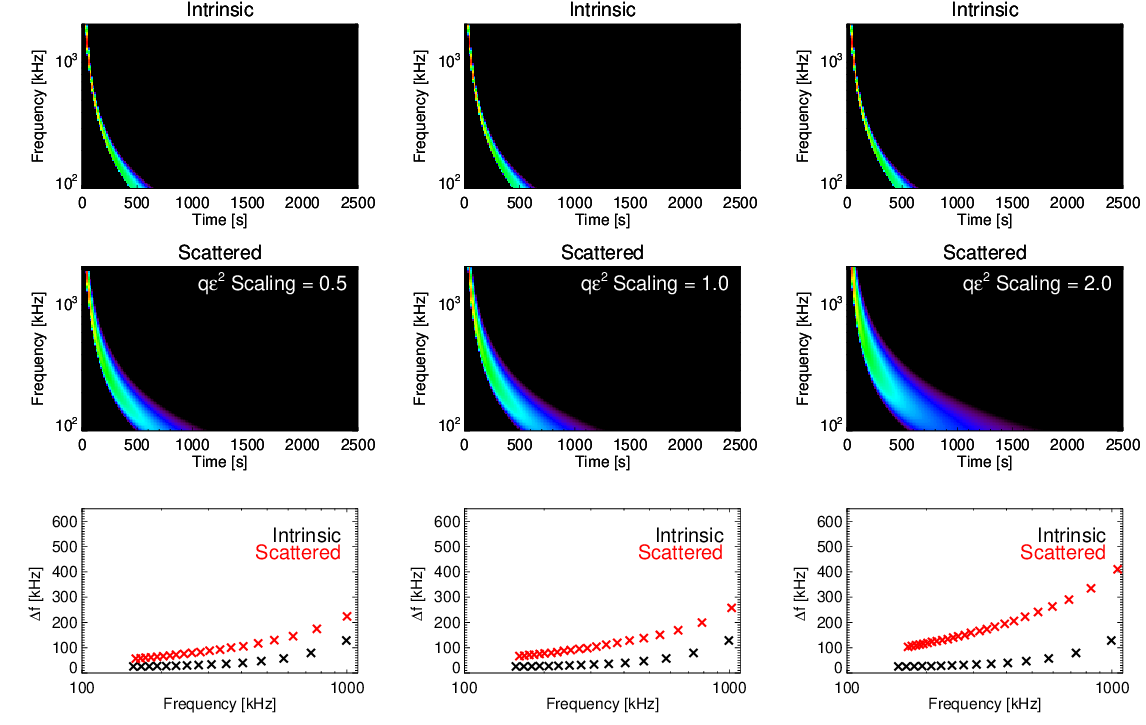}
    \caption{Analytical Type III bursts and their associated instantaneous bandwidths. The intrinsic dynamic spectra in the top row were generated using equation \ref{eq:analytical_typeIII} where $d=2\times10^{10}$~cm, $v=0.3c$, and $L=10$. Each column shows the same intrinsic burst, and a scattered burst using different turbulence scaling factors of $\overline{q\epsilon^2}$ from 0.5 (left column; lower scattering rate), 1.0 (centre column, nominal scattering rate), and 2.0 (right column, higher scattering rate).} The anisotropy level is kept at $\alpha=0.25$ for all examples. The lower panels show the measured values of $\Delta{f}$ against the peak frequency of each $I(f)$ profile for both the intrinsic burst (black) and scattered burst (red).
    \label{fig:beam_turbScaling_comparison}
\end{figure}
Type III bursts tend to have a maximum intensity near $\sim2$~MHz with \textit{individual} bursts typically presenting a drop of approximately 1--2 orders of magnitude towards 0.1~MHz \citep[e.g.][]{2014SoPh..289.3121K, 2022ApJ...924...58S}. We model this behaviour from 2 to 0.1~MHz through the weighting term $w_i(f(r)) = (f_\mathrm{MHz}(r)/2\,\mathrm{MHz})^{0.5}$. The intrinsic analytical type III bursts are shown in the upper row of Figures (\ref{fig:beam_velocity_comparison})--(\ref{fig:beam_Lexp_comparison}).

To emulate scatter broadening of the time profiles, we take the initial light-curves of the intrinsic type III bursts at a given frequency, and convolve with a profile representing the contribution due to scattering. We model the scattering function $I_\mathrm{scat}(t)$ using:
\begin{equation}\label{eq:It}
    I_\mathrm{scat}(t) = \left[I_0\exp\left(-\frac{\tau_1}{t-t_0} - \frac{t-t_0}{\tau_2}\right) + C \right] \times H(t,t_o).
\end{equation}
Here, $t_0$ is a time offset for the activation of the Heaviside function $H(t,t_0)$, as described and previously used to fit kilometric burst time profiles by \cite{2024A&A...687L..12C}. To specify the required decay time at a given frequency, we take the $1/e$ decay times $t_d$ from fundamental radio-wave propagation simulations used in \cite{2023ApJ...956..112K} for anisotropy value of $\alpha=0.25$ (see their Figure 3). Their heliospheric turbulence model is defined by the spectrum-weighted mean wavenumber $\overline{q\epsilon^2}$, which as a function of distance is given by $\overline{q\epsilon^2}(r)=2\times10^3\alpha R_\odot\left(1-r/R_\odot\right)^{-2.7}\left(r/R_\odot\right)^{-0.7}$, where $\epsilon=\langle\delta n^2/n^2\rangle$ is the density fluctuation variance. This is proportional to the scattering rate of the radio-waves \citep{2019ApJ...884..122K}. A spread in the scattering rate between a scaling factor of 0.5--2 was found to explain the majority of solar and extra-solar radio data \citep{2023ApJ...956..112K}. Since the simulations are run at discrete frequencies, we fit the decay times $t_d(f)$ with the power law function of the form $t_d(f)=\beta f^\eta$ across the frequency range  0.1--2~MHz, giving $\beta=5.81\times10^5$ and $\eta=-1.37$. We note that this is steeper than the observed type III burst decay times that follow $f^{-1}$ over broad frequency range \citep[e.g.][]{2023ApJ...956..112K, 2024ApJ...974L..18V}; however, as the simulations by \cite{2023ApJ...956..112K} show, the scattering contribution to observed decay times steepens below $\sim10$~MHz, approximately where the solar wind becomes supersonic and where turbulence level (and proportionally the scattering rate) in the turbulence model of $\overline{q\epsilon^2}$ has a broad maximum. In equation \ref{eq:It}, $\tau_1$ and $\tau_2$ do not directly correlate to the rise and decay times---we find setting $\tau_1=\tau_2=0.62t_\mathrm{d}$ generates asymmetrical functions that maintain the required decay time. The convolution $I_\mathrm{int}(t) * I_\mathrm{scat}(t)$ for a given frequency channel is then given by
\begin{equation}
    I_\mathrm{int}(t) * I_\mathrm{scat}(t) = \int_{-\infty}^{\infty}{ I_\mathrm{int}(t-t')I_\mathrm{scat}(t')\,\mathrm{d}t'}.
\end{equation}
The resulting convolved dynamic spectra are shown in the middle panels of Figures (\ref{fig:beam_velocity_comparison})--(\ref{fig:beam_Lexp_comparison}). Through the remainder of this work, we refer to the bursts before and after convolution as intrinsic and scattered, respectively. The scattered burst profiles are noticeably broader in time and frequency, and the timing of the peak intensity is shifted to later times, similar to simulations by \cite{2023MNRAS.520.3117C}..

\begin{figure}[htb!]
    \centering
    \includegraphics[width=.5\linewidth]{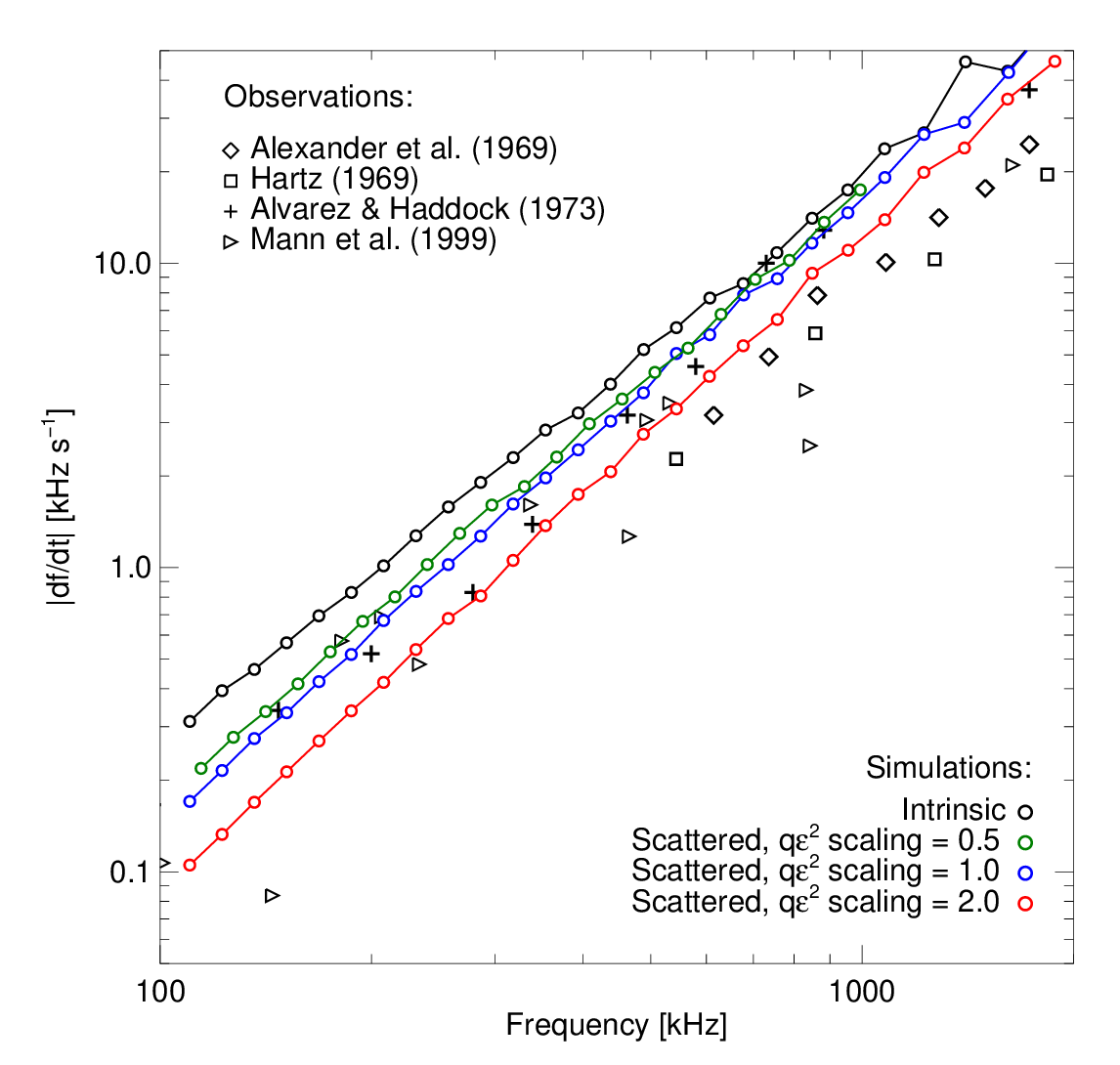}
    \caption{Frequency drift rate of the analytical type III bursts shown in Figure \ref{fig:beam_turbScaling_comparison}. The open black circles show the drift-rate before convolution with a scatter function, and the coloured open circles show the drift rate after convolution for different turbulence scaling factors. The remaining black data points show observed measurements from \cite{1969SoPh....8..388A, 1969P&SS...17..267H, 1973SoPh...29..197A, 1999A&A...348..614M}.}
    \label{fig:sim_dfdt_before_after}
\end{figure}

\begin{figure}[htb!]
    \centering
    \includegraphics[width=1\linewidth]{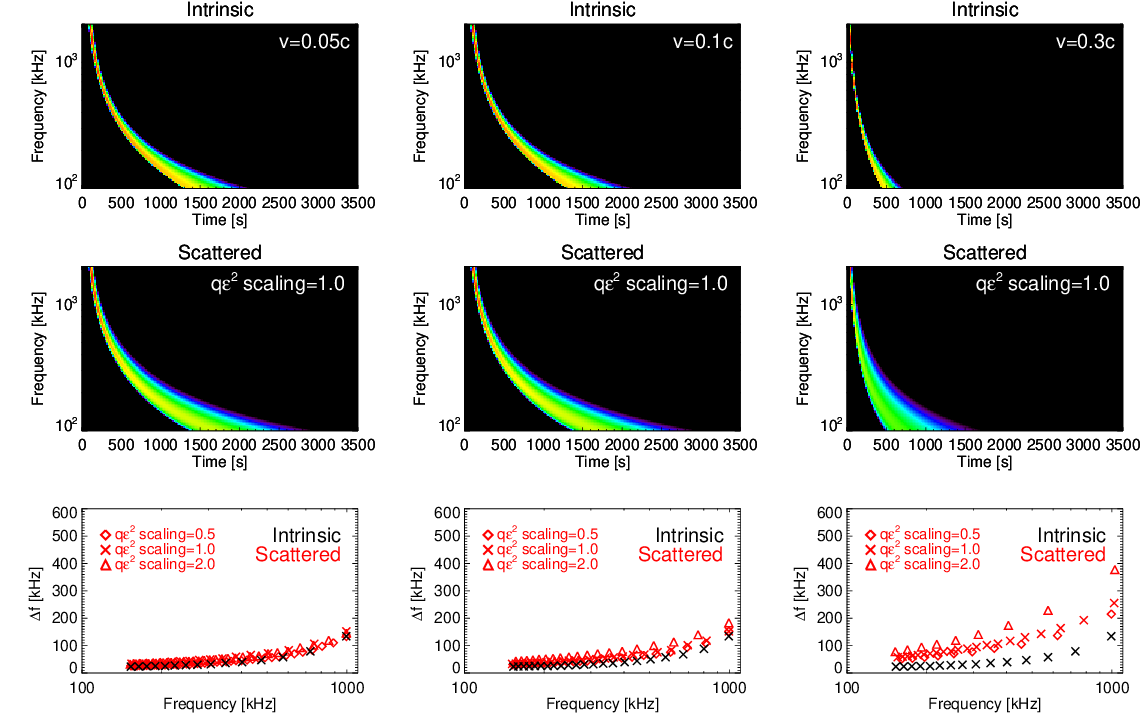}
    \caption{As in Figure \ref{fig:beam_turbScaling_comparison}, using a beam size of $d=2\times10^{10}$~cm and characteristic expansion scale $L=10$. Each column shows a different beam velocity $v$ from $0.05c$, $0.1c$, and $0.3c$. Also presented are the bandwidths of scattered bursts where $\overline{q\epsilon^2}$ scaling is 0.5 and 2.0.}
    \label{fig:beam_velocity_comparison}
\end{figure}

\subsubsection{Turbulence Level}

Evidently, different levels of turbulence will affect the broadening of radio time profiles. In Figure \ref{fig:beam_turbScaling_comparison}, we show how this effect changes the observed bandwidth of type III burst emission. 
In each panel of the upper row, the intrinsic type III burst emission is for fixed beam parameters of $d=2\times10^{10}$~cm, $L=10$~R$_\odot$, and $v=0.3c$. The middle row shows the scattered type III bursts -- i.e. the convolution of the intrinsic bursts in the upper panels using scattering kernels whose decay times correspond to turbulence scaling factors 
of $0.5$--$2$, which is the range of density fluctuations observed in the heliosphere \citep{2023ApJ...956..112K}. 
The lower row shows the bandwidths of both the intrinsic and scattered bursts from the upper and middle panels. In each example, the scattered burst has a broader bandwidth than the intrinsic burst. Increasing turbulence produces broader time profiles, resulting in larger type III burst bandwidths. 
For example, for the lowest turbulence level considered (scaling factor of 0.5), the smallest bandwidth measured is $\Delta{f}\sim100$~kHz, whereas for double the turbulence level (scaling factor of 2), the smallest bandwidth is almost 200~kHz. In addition, the frequency of peak intensity in the $I(f)$ profile is increased to higher frequencies for the scattered bursts. Further, scatter broadening modifies the rate at which the bandwidth $\Delta{f}$ changes with peak frequency $f$. For the weakest turbulence level considered, the frequency dependence is similar to that of the intrinsic burst, but becomes progressively steeper as the turbulence level increases. At higher frequencies, the intrinsic type III time profiles are comparatively short, so the scattering-induced decay time becomes a larger fraction of the total burst duration, leading to a larger increase in the measured bandwidth. These simulations demonstrate that radio-wave scattering can make a substantial contribution to the observed bandwidth, particularly at higher frequencies and under stronger levels of turbulence, justifying the inclusion of the $\Delta{f}_\mathrm{scat}$ term in Equation \ref{eq:df}.

Scatter broadening also causes a reduction in the frequency-drift rate of the burst. Figure \ref{fig:sim_dfdt_before_after} shows the frequency drift-rates of the simulated type III bursts from Figure \ref{fig:beam_turbScaling_comparison}, comparing the intrinsic and scattered bursts. The scattering-induced temporal broadening delays the time of maximum intensity, reducing the measured drift rate. This effect becomes progressively stronger with increasing turbulence, as the longer scattering decay times produce greater temporal broadening. For the strongest turbulence level considered, with a $\overline{q\epsilon^2}$ scaling factor of 2, the drift rate is reduced by a factor of $\sim3$ near 0.1~MHz. Thus, in addition to broadening the instantaneous bandwidth, radio-wave scattering can substantially modify the observed frequency drift-rate of type III bursts.

\subsubsection{Beam Velocity}

A slower propagating electron beam produces a smaller frequency drift rate in the dynamic spectrum of a type III burst. In Figure \ref{fig:beam_velocity_comparison}, we explore beam velocities ranging from $v=0.05c$ to $v=0.3c$. For each case, the scattered burst is calculated assuming a turbulence scaling factor of 1 and anisotropy of 0.25. Comparing the intrinsic bursts, varying the beam velocity changes only the frequency drift rate, while the intrinsic bandwidth remains essentially unchanged because the emitting spatial length is fixed. In contrast, the scattered bursts exhibit progressively larger bandwidths with increasing beam velocity. For $v=0.05c$, the bandwidth of the scattered burst is nearly identical to that of the intrinsic burst, whilst at $v=0.1c$, a modest increase is observed, increasing substantially for $v=0.3c$. This behaviour mirrors the frequency dependence found for increasing turbulence levels in Figure~\ref{fig:beam_turbScaling_comparison}. As the influence of scattering grows, whether through stronger turbulence or faster beam velocities, the enhancement of the measured bandwidth becomes increasingly pronounced towards higher frequencies. For weak scattering influence, either due to low turbulence levels or slow beam velocities, the intrinsic bandwidth dominates and little frequency dependence is observed. This demonstrates that radio-wave scattering has little influence on the measured bandwidth for slowly propagating electron beams $(v\lesssim0.1c)$, where the bandwidth is dominated by the intrinsic emitting spatial extent. As the beam velocity increases $(v\gtrsim0.1c)$, scattering contributes an increasingly significant fraction of the observed bandwidth, such that the measured bandwidth no longer reflects only the intrinsic emission region.

\subsubsection{Emitting Spatial Length}

\begin{figure}[htb!]
    \centering
    \includegraphics[width=1\linewidth]{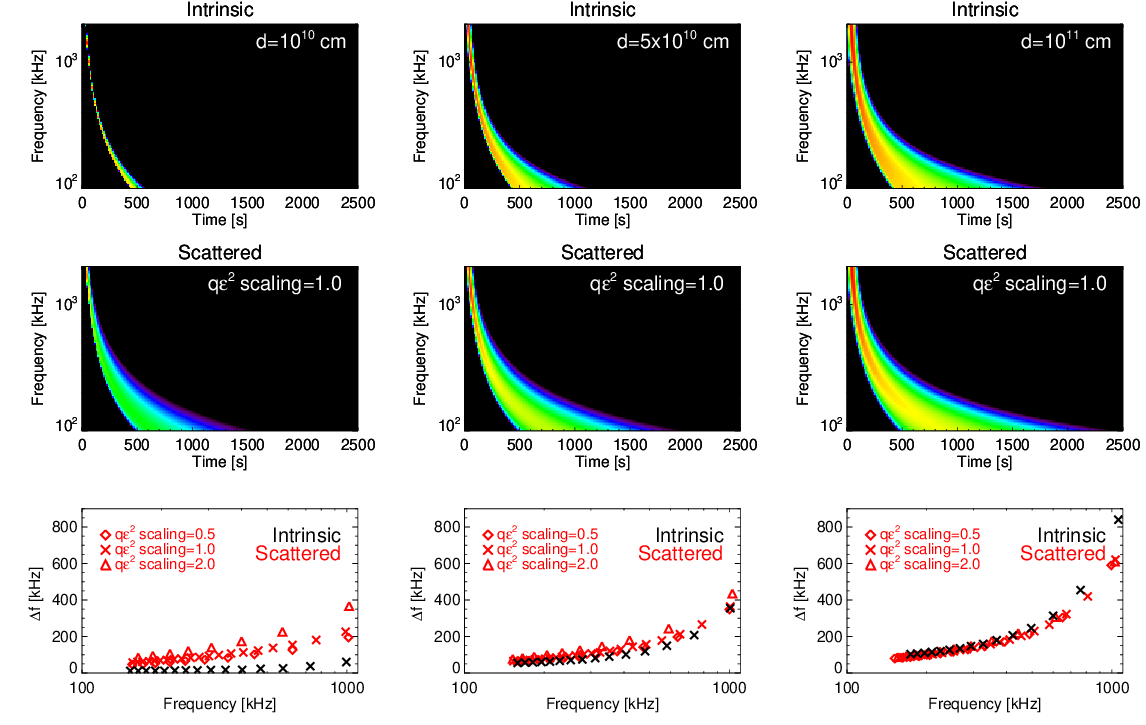}
    \caption{As in Figure \ref{fig:beam_velocity_comparison}, using a beam velocity of $v=0.3c$ and characteristic expansion scale $L=10$. Across each column, the beam size $d$ is varied with lengths of $10^{10}$~cm, $5\times10^{10}$~cm, and $10^{11}$~cm. Also presented are the bandwidths of scattered bursts where $\overline{q\epsilon^2}$ scaling is 0.5 and 2.0.}
    \label{fig:beam_size_comparison}
\end{figure}

Figure \ref{fig:beam_size_comparison} explores different emitting spatial widths $d=[10^{10}, 5\times10^{10}, 10^{11}]$~cm. Increasing the emitting spatial length broadens the intrinsic type III burst, particularly at higher frequencies where the bandwidth increases systematically with $d$. For the smallest emitting region ($d=10^{10}$~cm), the intrinsic bandwidth is much smaller than the scattered bandwidth, indicating that the measured bandwidth is dominated by radio-wave propagation. Increasing the emitting spatial length to $5\times10^{10}$~cm produces little change in the scattered bandwidth itself, but the intrinsic bandwidth grows substantially and becomes comparable to the scattered bandwidth, so scattering can no longer be considered the sole dominant contribution to the observed broadening. When $d=10^{11}$~cm, the intrinsic and scattered bandwidths are nearly indistinguishable, indicating the transition to a regime where intrinsic source effects dominate the measured bandwidth. This interpretation is consistent with the simple quadrature approximation of Equation~(\ref{eq:df}). Increasing the intrinsic bandwidth has little effect on the observed bandwidth while $\Delta f_\mathrm{int}\ll\Delta f_\mathrm{scat}$, but becomes increasingly important once the intrinsic and scattering contributions are comparable.

\subsubsection{Emitting Region Expansion}

\begin{figure}[htb!]
    \centering
    \includegraphics[width=1\linewidth]{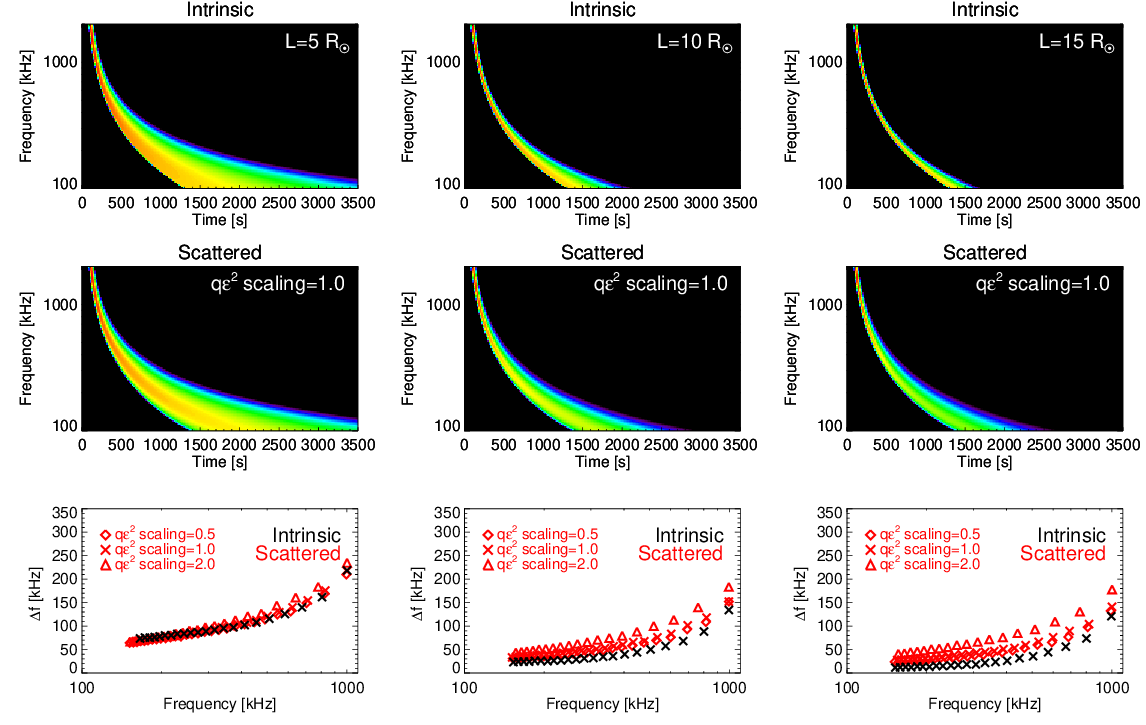}
    \caption{As in Figure \ref{fig:beam_velocity_comparison}, using a beam velocity of $v=0.3c$ and length $d=2\times10^{10}$~cm. Across each column, the characteristic expansion scale $L$ is varied from 5~R$_\odot$, 10~R$_\odot$, and 15~R$_\odot$. Also presented are the bandwidths of scattered bursts where $\overline{q\epsilon^2}$ scaling is 0.5 and 2.0.}
    \label{fig:beam_Lexp_comparison}
\end{figure}

Expansion of the emitting region is controlled by the parameter $L$. Figure \ref{fig:beam_Lexp_comparison} explores values of $L=[5, 10, 15]$~R$_\odot$. Smaller values of $L$ produce a more rapidly expanding emitting region, resulting in broader intrinsic emission profiles in both time and frequency. For $L=[10, 15]$~R$_\odot$, the intrinsic bandwidth is sufficiently narrow that the observed bandwidth is primarily determined by scattering. Consequently, the intrinsic bandwidth has little effect on the scattered burst bandwidth, which remains relatively unchanged across these cases. In contrast, for $L=5$~R$_\odot$, the intrinsic bandwidth becomes comparable to the scattered bandwidth. In this regime, scattering has a weaker influence on the overall bandwidth, similar to the behaviour observed for lower beam velocities and larger emitting spatial lengths. This demonstrates that scattering only significantly modifies the observed bandwidth when the intrinsic bandwidth is sufficiently narrow; when the intrinsic emission is already broad, scattering produces comparatively little additional broadening.

\section{Discussion and Conclusion}\label{sec:discussion_conclusion}

The instantaneous bandwidth of type III radio bursts has generally been interpreted as a measure of the spatial extent of the emitting exciter. This interpretation implicitly assumes that the observed radio emission primarily reflects the intrinsic plasma emission process. By combining statistical observations of kilometric type III bursts with numerical simulations of electron beam evolution and analytical models of radio-wave scattering, the present work revisits this interpretation. Our results indicate that the observed bandwidth is best understood as the combination of an intrinsic contribution associated with the plasma emission process and an additional contribution arising from radio-wave propagation induced temporal broadening.

Previous numerical studies of plasma emission have demonstrated that the intrinsic radio bandwidth follows the spatial extent of Langmuir wave generation rather than the full spatial extent of the electron beam \citep{2014A&A...572A.111R, 2018A&A...614A..69R}. Our numerical simulations reproduce this behaviour, with the simulated radio bandwidth closely following the Langmuir wave energy density by construction while remaining narrower than the electron beam itself. When compared with the statistical bandwidth measurements presented here, however, the intrinsic bandwidth produced by the plasma emission process occupies only the lower boundary of the observed distribution. Moreover, the radio-wave transport simulations demonstrate how an additional contribution from radio-wave scattering can change the intrinsic and observed bandwidths.

The required additional broadening can arise from radio-wave scattering, which modifies the temporal profile of the emission without substantially changing its frequency. 
Although scattering conserves frequency of radio 
waves \citep[excluding inelastic scattering effects, see][]{2024ApJ...968...72A}, it substantially broadens burst time profiles through temporal broadening at a given frequency. 
As the type III bursts drift rapidly through frequency space, this temporal broadening causes emission from neighbouring frequencies to overlap in time, increasing the apparent instantaneous bandwidth. 
The importance of this effect depends on the relative magnitude of the intrinsic and propagation-induced contributions. 
Stronger density turbulence produces larger temporal broadening and correspondingly broader observed bandwidths, 
while faster electron beams are affected more strongly because their intrinsically shorter burst durations allow scattering to dominate a larger fraction of the observed profile. 
Conversely, for slowly propagating beams or intrinsically broad emission regions, the measured bandwidth remains largely controlled by the intrinsic plasma emission process.

The statistical measurements of Type III bursts presented here extend previous bandwidth measurements into the kilometric regime and show how they remain broadly consistent with measurements made at decametric and metric frequencies. The approximately continuous increase of bandwidth with frequency suggests that the same underlying physical processes operate over a wide range of heliocentric distances. The simultaneous observations from multiple spacecraft provide an additional observational constraint. Despite the largest heliolongitude separations of approximately $147\arcdeg$ and $166\arcdeg$ across two events, all spacecraft measure remarkably similar bandwidths for the same type III burst. This similarity could indicate either that the bandwidth is dominated by the intrinsic emitting region (which does not depend on observer position), or that the scattering contribution to the time profile itself has little angular dependence. The latter is consistent with \cite{2024A&A...687L..12C}, who find no dependence of burst decay times on heliospheric longitude despite decay times being largely governed by radio-wave propagation. By contrast, \cite{2025ApJ...989..118A} find that the inferred velocity and acceleration of type III emitters, derived from frequency drift-rate, does depend on which spacecraft is used to view the burst - indicating that not every diagnostic is independent of viewing geometry. 
Although the present observations cannot uniquely separate intrinsic and propagation contributions, they suggest that the scattering-induced component of the bandwidth is not strongly dependent on viewing geometry for the events considered here. Future comparisons with radio-wave propagation simulations and larger samples of multi-spacecraft observations should provide further constraints on this behaviour.

More broadly, these results demonstrate that instantaneous bandwidth should not be regarded as a purely source-related diagnostic. Alongside apparent source size, burst decay time, and frequency drift rate, the observed bandwidth is significantly modified by radio-wave propagation. Consequently, estimates of emitting source sizes or electron beam properties based solely on measured bandwidths may systematically overestimate the intrinsic emission region if propagation effects are neglected. 
Conversely, because the observed bandwidth depends on both the intrinsic plasma emission process and the strength of radio-wave scattering, it also offers the possibility of diagnosing the turbulent plasma through which the emission propagates when interpreted together with independent constraints on the electron beam properties. Furthermore, polarisation measurements may provide an additional diagnostic of radio-wave propagation effects on the observed bandwidth. 
The degree of circular polarisation of type III bursts can decrease during the burst decay, reflecting depolarisation during propagation through the turbulent corona \citep[e.g.][]{1980A&A....88..203D, 2017NatCo...8.1515K, 2020SoPh..295...51R}. Frequency- and time-resolved polarisation measurements could therefore help identify intervals in which the emission remains strongly polarised, providing an opportunity to assess the bandwidth before substantial depolarisation occurs. 
This could help constrain the contribution of propagation effects to the observed bandwidth and improve estimates of the intrinsic component.

The models presented here necessarily adopt several simplifying assumptions. The numerical simulations treat electron beam transport and Langmuir wave evolution in one spatial dimension, while the analytical model represents the intrinsic radio emission through a prescribed asymmetric burst profile rather than modelling the complete plasma emission process. 
Furthermore, the scattering contribution is parameterised using decay times derived from radio-wave propagation simulations and therefore samples a limited, yet broadly realistic, range of turbulence conditions \citep{2023ApJ...956..112K}. These assumptions influence the quantitative contribution of scattering to the measured bandwidth but do not alter the principal conclusion that propagation-induced temporal broadening can substantially increase the observed instantaneous bandwidth. Moreover, the majority of the results here assume fundamental plasma emission. Although the qualitative effect of propagation-induced temporal broadening on the instantaneous bandwidth is not specific to the emission mode, the quantitative results can differ for harmonic emission because of differences in the emission location and propagation conditions (Appendix section \ref{sec:harmonic_emission}).

Overall, this work supports a revised interpretation of type III burst bandwidths. The intrinsic bandwidth is determined by the plasma emission process and reflects the spatial extent of Langmuir wave generation, consistent with previous numerical studies. However, this intrinsic contribution alone cannot explain the full range of observed bandwidths. Radio-wave scattering introduces an additional propagation-induced broadening that can become comparable to, or even dominate, the intrinsic contribution depending on the turbulence level and electron beam properties. Consequently, the instantaneous bandwidth of type III radio bursts should be interpreted as the combined result of intrinsic plasma emission and radio-wave propagation, rather than as a direct measure of the spatial extent of the emitting region alone.

\begin{acknowledgments}
EPK and DLC acknowledge financial support from the STFC/UKRI grant ST/Y001834/1; EPK is also supported by the Leverhulme Trust (Research Fellowship RF-2025-357). The authors thank the PSP/RFS, SolO/RPW, and STEREO/WAVES teams for making the data available. Solar Orbiter \citep{2020A&A...642A...1M} is a mission of international cooperation between ESA and NASA, operated by ESA. The FIELDS experiment on the Parker Solar Probe spacecraft \citep{2016SSRv..204....7F} was designed and developed under NASA contract NNN06AA01C. This research has made use of the Astrophysics Data System, funded by NASA under Cooperative Agreement 80NSSC21M00561.
\end{acknowledgments}

\appendix
\counterwithin{figure}{section}
\renewcommand\thefigure{\thesection\arabic{figure}}

\section{Bandwidth of Scattered Harmonic Emission}\label{sec:harmonic_emission}

\begin{figure}[b!]
    \centering
    \includegraphics[width=1\linewidth]{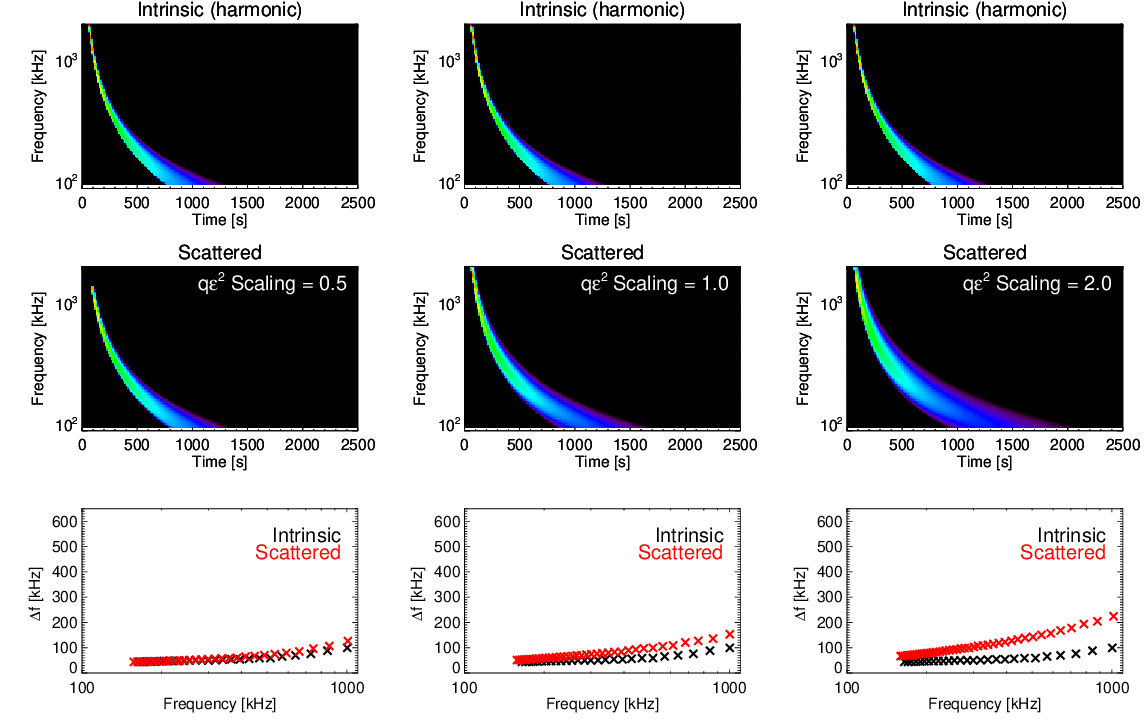}
    \caption{As in Figure \ref{fig:beam_turbScaling_comparison}, for harmonic radio emission.}
    \label{fig:beam_turbScaling_comparison_harmonic}
\end{figure}

Figure \ref{fig:beam_turbScaling_comparison_harmonic} presents analytical simulations where the emission mode is considered at the harmonic level. For a given observing frequency, harmonic emission originates at a greater heliocentric distance than fundamental emission, where the plasma density is lower. Consequently, for the same beam velocity, density gradient, and observing frequency, harmonic emission exhibits a slower intrinsic frequency drift rate than the fundamental, as the beam takes longer to traverse a given frequency interval. Here, the time profiles representing the contribution due to scattering have a decay time matching the harmonic simulations of \cite{2023ApJ...956..112K}. For comparison, we consider the decay times for the same anisotropy of $\alpha=0.25$ as used for the fundamental equivalent. The degree of bandwidth broadening that is observable in dynamic spectra due to temporal broadening of the burst profile depends on the intrinsic frequency drift rate, as demonstrated in Figure \ref{fig:beam_velocity_comparison}. Consequently, the lower intrinsic frequency drift rate of harmonic emission at a given frequency compared to the fundamental affects the extent of scattering-induced bandwidth broadening, without implying weaker scattering during propagation.

\clearpage
\bibliographystyle{aasjournal}
\bibliography{refs}

\end{document}